\documentclass[
aip,
amsmath,amssymb,
reprint
]{revtex4-2}

\usepackage{comment}

\usepackage{graphicx}
\usepackage{dcolumn}
\usepackage{amsfonts}
\usepackage{dsfont}
\usepackage{mathrsfs}
\usepackage{enumitem}
\usepackage{footnote}
\usepackage{booktabs}
\usepackage{braket}
\usepackage[overload]{empheq}
\usepackage{titlesec}
\usepackage{array}
\usepackage{upgreek}
\usepackage[usenames, dvipsnames]{xcolor}
\usepackage{siunitx}
\usepackage{tabularx}
\usepackage{cases}
\usepackage{epstopdf}
\usepackage{framed}
\usepackage{xspace}
\usepackage{pifont}
\usepackage{hhline}
\usepackage{multirow}
\usepackage{mwe}
\usepackage[utf8]{inputenc}
\usepackage[T1]{fontenc}
\usepackage{bm}
\usepackage{hyperref}
\usepackage{ulem}

\begin{document}
	
	\title{Accelerating charging dynamics of electric double-layer capacitors}

	\author{Megh Dutta}
	\affiliation{ 
		Physicochimie des \'Electrolytes et Nanosyst\`emes Interfaciaux, Sorbonne Universit\'e, CNRS, Paris, 75005, France
	}
	
	\author{Ivan Palaia}
	\affiliation{
		Department of Physics, King’s College London, WC2R 2LS, United Kingdom 
	}
	
	\author{Emmanuel Trizac}
	\affiliation{
		Universit\'e Paris-Saclay, CNRS, LPTMS, 91405 Orsay, France 
	}
	\affiliation{
		\'Ecole Normale Sup\'erieure de Lyon, F-69342, Lyon, France}
	\author{Benjamin Rotenberg}
	\email[]{benjamin.rotenberg@sorbonne-universite.fr}
	\affiliation{ 
		Physicochimie des \'Electrolytes et Nanosyst\`emes Interfaciaux, Sorbonne Universit\'e, CNRS, Paris, 75005, France
	}
	\affiliation{Réseau sur le Stockage Electrochimique de l’Energie (RS2E), FR CNRS 3459, 80039 Amiens Cedex, France}
	
	\date{\today}
	
	\begin{abstract}
		Electric double-layer capacitors (EDLCs), consisting of an ionic fluid between two metallic electrodes, are electrochemical energy storage devices complementary to batteries, allowing for a faster charge/discharge. The charging dynamics in response to a voltage step features a variety of regimes and relaxation timescales, depending on the applied voltage and the various lengths characterizing the system, most importantly the inter-electrode distance and the Debye length over which electrostatic effects are screened in the electrolyte. Inspired by recent works on ``shortcut to adiabaticity'' in colloidal systems, here we investigate the possibility to control the charge and discharge of planar EDLCs using time-dependent voltages. Specifically, our aim is to achieve a full charge or discharge within a finite time shorter than their intrinsic relaxation timescales. Within the Poisson-Nernst-Planck model and the small-voltage regime, we derive time-dependent protocols that can eliminate an arbitrary number of relaxation modes. This permits to approach the equilibrium charged state within a finite time, that can be in practice an order of magnitude faster than the natural equilibration time. We illustrate the relevance and efficacy of the method on polynomial drivings and show that the surface charge density, charge-density profiles, and global deviation from equilibrium (quantified by a Kullback–Leibler-like divergence) can all be significantly accelerated, even for driving times comparable to or shorter than the natural $RC$ time of the system.
	\end{abstract}
	
	\maketitle

	\section{Introduction}
	
	Ionic fluids, formed by ions usually dissolved in a polar solvent, exhibit interesting properties near surfaces or under confinement. If the surface is electrically charged, an electrolytic double-layer (EDL) forms at the interface, with a local net charge in the liquid arising from an imbalance in counter-ions (charge of opposite sign compared to the surface) and co-ions (same sign)~\cite{parsons_electrical_1990}. The surface charge can be induced by applying a voltage between two conductors separated by an electrolyte, resulting in an electric double layer capacitor (EDLC)~\cite{simon_perspectives_2020}. In contrast to standard capacitors, which develop a polarization throughout the polarizable insulating material separating the electrodes, in these systems, at equilibrium the electric field is screened at the interface so that the rest of the liquid is not polarized~\cite{Ji2014, Frivaldsky2018, TORKI2023106330}. In contrast to batteries, their charging does not involve electron transfer reactions at the interface, which enables faster charge/discharge (at the price of lower energy density) and their stability over many cycles, fostering their use in a wealth of applications~\cite{Miller_2008, DURGANJALI2022106140}. 
	
	Given the relevance of charging dynamics in EDLCs, several strategies have been proposed to address different aspects from the theoretical point of view, from analytical theories based on continuous descriptions to atomistic simulations via mesoscopic models, in order to identify generic scaling laws or to understand specific molecular aspects of the electrode/electrolyte interface or the transport of ions inside porous electrodes~\cite{pean_dynamics_2014, pean_confinement_2015, pean_multi-scale_2016, noh_understanding_2019, jeanmairet_microscopic_2022, waysenson_electrode_2025, Asta2019JCP, lin_microscopic_2022, Ma2022, wu_understanding_2022, kondrat_theory_2023}. Experimentally, the dynamics of charging is investigated using electrochemical impedance spectroscopy~\cite{wang_electrochemical_2021, vivier_impedance_2022,doi:10.1021/acs.jpcc.8b05559,doi:10.1021/jp5025476}, which characterizes the frequency-dependent response of the systems to small voltages. This linear response can be predicted based on various models of the dynamics, at the continuous level~\cite{PhysRev.92.4,barbero_impedance_2005, barbero_role_2005, barbero_theory_2008, barbero_theoretical_2017} or from mesoscopic and molecular simulations~\cite{pireddu_frequency-dependent_2023, pireddu_impedance_2024}. This allows to link transport through the EDLC with the equivalent circuit models used to interpret the experiments, such as the simple $RC$ circuit to capture the accumulation of charge at the interface and the dissipation due to ionic transport through the electrolyte (see \textit{e.g.} Ref.~\citenum{barnaveli_asymmetric_2024} for a recent example with a slightly more elaborate circuit in the thin EDL limit).

	Following the pioneering investigation of Bazant \textit{et al.}~\cite{Bazant2004}, theoretical work based on the Poisson-Nernst-Planck model to describe the dynamics of ions between the electrodes, or its extensions, allowed to identify various regimes depending on the applied voltage and the competition between various length scales (see Fig.~\ref{fig:EDLC}): the inter-electrode distance $2L$, the Debye length $\lambda_D$ characterizing the screening of electrostatic interactions between the ions and the extent of the EDL, or the Bjerrum length beyond which the electrostatic energy between two unit charges is smaller than the thermal energy~\cite{Bazant2004, Beunis2008, PhysRevLett.115.106101, Janssen2018, Palaia2019, palaia_charging_2025, palaia_pnp_2025, fertigcharging}. These works highlighted in particular the importance of the timescale $L\lambda_D/D$, with $D$ a typical diffusion coefficient of the ions, corresponding to the $RC$ charging time in the thin EDL limit ($\lambda_D\ll L)$.
	
	Since the key advantage of EDLCs compared to batteries is their charging dynamics, a natural question arises: Is it possible to reduce their charging and discharging time using time-dependent voltages, \textit{i.e.} achieve a full charge or discharge within a finite time shorter than their intrinsic relaxation timescales? This could be useful not only in the context of energy storage/delivery, but also for other applications. This question has already fostered several attempts. For example, Breitsprecher \textit{et al.} showed using molecular simulations that the discharge of a capacitor with nanoporous electrodes can be accelerated using a voltage inversion followed by a linear decay, compared to simply switching the voltage off~\cite{Breitsprecher2020,Breitsprecher2018}, while Lian \textit{et al.} used mesoscopic modelling of EDLCs with porous electrodes to discuss the necessary trade-off between capacitance and charging time~\cite{Lian2020}. In this context, another relevant aspect is the energetics and how quantities such as the necessary work or the dissipated heat depends on the charging protocol~\cite{PhysRevLett.118.096001, C5EE01192B, PhysRevLett.113.268501, C3CP44612C}.
	
	In the present work, we address this question more systematically. To this end, we build upon recent developments in other contexts, where protocols have been derived to enable faster transitions between two equilibrium states. Inspired from the notion of ``shortcut to adiabaticity'' in quantum systems, several strategies have been developed for stochastic systems where the dynamics depends on thermal fluctuations, with successful applications in soft matter~\cite{Martinez2016, bayati_diffusiophoresis_2021, plata_taming_2021, raynal_shortcuts_2023, RevModPhys.91.045001,guery-odelin_2023}. Here, we do not follow standard approaches of optimal control theory relying on the minimization of a predefined cost functional. Namely, using linear response theory in the Laplace domain, we design an input (the driving voltage) that suppresses the slowest relaxation modes from the transfer function, so that they are absent from the outputs (the ion distribution and corresponding electrode surface charge density), in order to approach the equilibrium charged state within a finite time.
	Section~\ref{sec:model} introduces the Poisson-Nernst-Planck model used to describe the ion dynamics in the EDLC. The response to a time-dependent charging protocol is then described in Section~\ref{sec:response}: the general solution is given in Section~\ref{sec:response:relaxationdynamics}, while the method to shortcut the charging time is introduced in Section~\ref{sec:response:shortcut}. Finally, we illustrate the relevance and efficacy of the method on polynomial drivings in Section~\ref{sec:response:polynomialdriving}.
	
	\section{Poisson-Nernst-Planck model of the ion dynamics}
	\label{sec:model}
	
	\begin{figure}
		\centering
		\includegraphics[width=0.9\columnwidth]{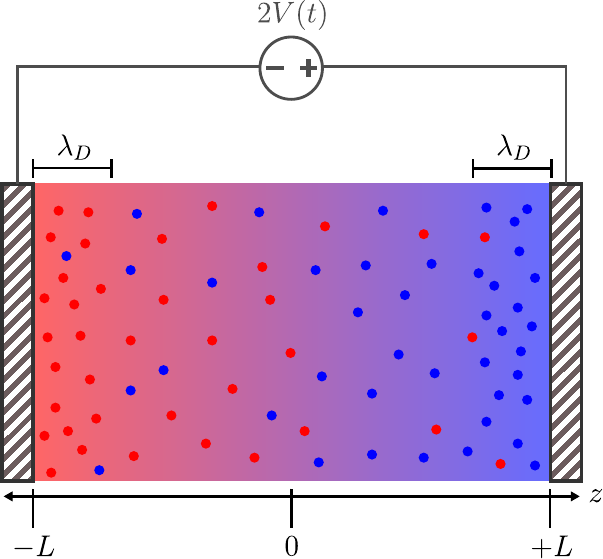}
		\caption{Diagram of an EDLC at equilibrium with the applied external potential difference. Within a mean-field description, cations (red) and anions (blue) are treated as a continuum, with values of the ionic densities  indicated by the color from red to blue. The Debye-length characterizing the electric double layer is indicated as $\lambda_D$.}
		\label{fig:EDLC}
	\end{figure}
	
	We consider the dynamics of cations and anions with charges $q_\pm e$ (where $e$ is the elementary charge) and diffusivities $D_\pm$ between two parallel, infinite metallic electrodes, separated by a distance $2L$ and submitted to a a time-dependent external potential difference $2V(t)$ (Fig. \ref{fig:EDLC}). For sufficiently low concentrations, one can treat electrostatic interactions between ions and with the electrodes at the mean-field level, neglect short-range interactions between ions, and only consider their exclusion from the solid walls via a no-flux boundary condition. We work in the canonical ensemble, with a fixed number of ion pairs per unit surface $2n_0 L$, where $n_0$ is the uniform density of the salt in the absence of voltage. For a symmetric electrolyte with $q_+=-q_-$, as we will consider below, the condition of overall electroneutrality imposes that the average cation and anion densities are $\overline{n}_\pm=n_0$.
	
	The ionic densities $n_\pm(z,t)$ are assumed to evolve according to the Poisson-Nernst-Planck (PNP) model, which combines the conservation equations
	\begin{equation}
		\frac{\partial n_\pm}{\partial t}(z,t)=-\frac{\partial j_\pm}{\partial z}(z,t)
		\label{eq:continuityeq}
	\end{equation}
	with an expression of the fluxes including drift due to the electric field and diffusion due to concentration gradients:
	\begin{equation}
		{j}_\pm(z,t)=-\beta D_\pm n_\pm(z,t)q_\pm e \frac{\partial\phi}{\partial z}(z,t) - D_\pm\frac{\partial n_\pm}{\partial z}(z,t)
		\; .
		\label{eq:current}
	\end{equation}
	Here, $\beta=1/k_BT$ with $k_B$ the Boltzmann constant and $T$ the temperature, and we have assumed that the diffusivity matrix is diagonal, \textit{i.e.} that the diffusion of one of the species is not influenced by any other species. In the following, we will also assume that $D_+=D_-=D$. Both equations result in the Nernst-Planck equation:
	\begin{equation}
		\frac{\partial n_\pm}{\partial t}(z,t)=D_\pm\frac{\partial}{\partial z} \left(n_\pm(z,t)\beta q_\pm e\frac{\partial\phi}{\partial z}(z,t)+\frac{\partial n_\pm}{\partial z}(z,t)\right)
		\label{eq:PNP}
	\end{equation}
	where the electric field $-\partial\phi/\partial z$ driving the ions includes that due to the external voltage and that arising from the charge distribution inside the liquid, $\rho=q_+ e n_+ + q_- e n_-$. The potential satisfies the Poisson equation:
	\begin{equation}
		\frac{\partial^2\phi}{\partial z^2}(z,t) =-\frac{\rho(z,t)}{\varepsilon_0\varepsilon_r}
		\label{eq:poisson}
	\end{equation}
	with $\varepsilon_0$ the vacuum permittivity and $\varepsilon_r$ the relative permittivity of the solvent. The PNP equations~\ref{eq:poisson} and~\ref{eq:PNP} must be solved with suitable boundary conditions. The potential at the electrodes satisfies $\phi(\pm L,t)=\pm V(t)$, the electric field at the interfaces is related to the surface charge density $\pm\sigma(t)$ of the electrodes, while the no-flux boundary conditions for the ions at $z_{b} =\pm L$ read: 
	\begin{align}
		- \beta q_\pm e n_\pm(z_b, t)  \frac{\partial\phi}{\partial z}(z_b, t) -\frac{\partial n_\pm}{\partial z}(z_b, t)&=0 \; .
		\label{eq:bcflux}
	\end{align}
	
	\begin{table}
		\begin{center}
			{\def\arraystretch{1.9}\tabcolsep=10pt
				\begin{tabular}{ l | c | c }
					Observable & Symbol & Unit \\ \hline
					\rule{-4pt}{5ex}
					Time & $t,\, \tau_i$ & \(\displaystyle \frac{L \lambda_D}{ D} \) \\
					Inverse time & $s,\, s_i$ & $\displaystyle \frac{ D}{L \lambda_D}$ \\
					Distance & $z$ & $L$ \\
					Volumic ion density & $n_\pm$ & $n_0$ \\
					Volumic charge density & $\rho$ & $2en_0$ \\
					Potential & $\phi,\, V$ & $\displaystyle \frac{1}{\beta e}$ \\
					Electric field & $E$ & $\displaystyle\frac{1}{\beta e L}$\\
					Surface charge density & $\sigma$ & $\displaystyle \frac{\varepsilon_0\varepsilon_r}{\beta eL}$\\
				\end{tabular}
			}
		\end{center}
		\caption{Units of the nondimensionalized physical quantities. The reference units are expressed as a function of the half-distance $L$ between electrodes, the inverse of the thermal energy $\beta=1/k_BT$, the elementary charge $e$ (we consider here a 1:1 electrolyte, \textit{i.e.} valencies $q_+=-q_-=1$), the salt concentration $n_0$, the common diffusion coefficient $D_\pm=D$ of ions, the vacuum permittivity $\varepsilon_0$ and solvent relative permittivity $\varepsilon_r$, and the resulting Debye screening length $\lambda_D$ (see Eq.~\ref{eq:deflambdaD}).
		}
		\label{tab:units}
	\end{table}
	
	Following Refs.~\citenum{palaia_charging_2025,palaia_pnp_2025}, we adimensionalize all physical quantities using the quantities summarized in Table~\ref{tab:units}. Importantly, the response of the system is fully determined by the ratio
	\begin{equation}
		\label{eq:defepsilon}
		\epsilon = \frac{\lambda_D}{L} 
	\end{equation}
	with the Debye screening length (for a 1:1 electrolyte with $q_+=-q_-=1$, as considered in the following)
	\begin{equation}
		\label{eq:deflambdaD}
		\lambda_D = \sqrt{\frac{\varepsilon_0\varepsilon_r}{2 n_0\beta e^2 }} \; ,
	\end{equation}
	and the reduced voltage
	\begin{equation}
		\label{eq:defv}
		v=\beta e V_0
	\end{equation}
	with a typical voltage $V_0$. In the present case, we use as reference the potential at steady-state $V_0=\lim_{t\to\infty}V(t)$.

	\section{Response to a time-dependent charging protocol}
	\label{sec:response}
	
	A full description of the various regimes (in terms of $\epsilon$ and $v$) for the response to a step potential $V(t)=v\,\Theta(t)$, with $\Theta$ the Heaviside step function, can be found in Refs.~\citenum{palaia_charging_2025,palaia_pnp_2025}. We address here a different problem, which is how to get as close as possible to the steady-state in a finite driving time $t_f$, using a time-dependent voltage protocol reaching the final value $v$ at $t_f$. To that end, we restrict ourselves to the linear response regime, which implies in particular $v\ll 1$. For convenience, we separate the steady-state and time-dependent parts of the potential as
	\begin{equation}
		\label{eq:defDeltaV}
		V(t)=v \,\Theta(t) +\Delta V(t)
	\end{equation}
	where $\Delta V(t)$ only varies between $t=0$ and $t_f$. The initial and final conditions $V(0)=0$ and $V(t_f)=v$ impose $\Delta V(0)=-v$ and $\Delta V(t_f)=0$.

	\subsection{Relaxation dynamics}
	\label{sec:response:relaxationdynamics}
	
	Using the adimensionalized quantities of Table~\ref{tab:units}, the linear response of the charge density profile $\rho(z,t)=[n_+(z,t)-n_-(z,t)]/2$ to a time-dependent driving voltage $V(t)$ can be conveniently found in the Laplace domain, introducing for time-dependent functions $f(t)$:
	\begin{equation}
		\label{eq:defLaplace}
		\widehat{f}(s) = \mathscr{L}\{f(t)\} = \int_0^\infty f(t) e^{-st} \, {\rm d}t \; .
	\end{equation}
	The solution for the charge density is~\cite{Palaia2019, palaia_charging_2025, palaia_pnp_2025}
	\begin{align}
		\label{eq:rhoLaplace}
		\widehat{\rho}(z,s)&=-\frac{\sinh\frac{z\sqrt{1+\epsilon s}}{\epsilon}}{\sinh\frac{\sqrt{1+\epsilon s}}{\epsilon}}H(s) \widehat{V}(s)
	\end{align}
	with the transfer function
	\begin{equation}
		\label{eq:defH}
		H(s)=\frac{1+\epsilon s}{1+s\sqrt{1+\epsilon s}\coth\frac{\sqrt{1+\epsilon s}}{\epsilon}}
		\; .
	\end{equation}
	The corresponding surface charge density, obtained from the field at the electrode surface, is
	\begin{align}
		\label{eq:sigmaLaplace}
		\widehat{\sigma}(s) & = -\frac{\sqrt{1+\epsilon s}}{\epsilon}\coth\frac{\sqrt{1+\epsilon s}}{\epsilon} H(s) \widehat{V}(s)
		\; .
	\end{align}
	
	We denote the complex poles of the transfer function $H(s)$ by $s_n$ and those of the driving potential $\widehat{V}(s)$ by $s_n^V$. The presence and nature of poles of the driving potential depends on the specific choice of $V(t)$. The complex poles of the transfer function can be determined numerically~\cite{Janssen2018, Palaia2019, palaia_charging_2025, palaia_pnp_2025}. They are all simple poles, real and negative and we order them by increasing absolute value ($|s_0|<|s_1|<\dots$). Their number is infinite and the corresponding timescales depend on $\epsilon=\lambda_D/L$. For strong EDL overlap ($\epsilon\to\infty$), the slowest mode corresponds to diffusion between the electrodes, with timescale $4L^2/(\pi^2D)$ (see Refs.~\citenum{Janssen2018, palaia_pnp_2025}). In the thin EDL limit ($\epsilon\to0$), the slowest two modes correspond to $RC$ charging time $L\lambda_D/D$ and Debye relaxation time $\lambda_D^2/D$ as also expected from frequency-domain analysis (see \textit{e.g.} Ref.~\citenum{chassagne_compensating_2016}). For $\epsilon=0.1$, considered in the illustrations below, the slowest mode corresponds to $s_0\approx-1.06$, \textit{i.e.} a relaxation time of $0.94$ in units of $L\lambda_D/D$, while the next ones correspond to $s_1\approx-12.05$, $s_2\approx-16.04$, $s_3\approx-22.00$ and $s_4\approx-29.92$, \textit{i.e.} faster and faster relaxation.
	
	In the general case, the residue theorem allows us to express the evolution of the charge density profile in the time domain, using Eq.~\ref{eq:rhoLaplace}, as
	\begin{align}
		\rho(z,t)&= \sum_{s\in\{s_n,s_n^V\}}\text{Res}_{\hat{\rho}(z,s)e^{st}}(s) 
		\, ,
		\label{eq:rhoresidues}
	\end{align}
	where the sum runs over poles of the transfer function and the driving potential. For the specific form of driving considered here, in the Laplace domain Eq.~\ref{eq:defDeltaV} leads to 
	\begin{equation}
		\widehat{V}(s)=\frac{v}{s}+\widehat{\Delta V}(s) \; .
	\end{equation}
	$\hat{\rho}(z,s)$ has a simple pole at $s=0$, coming not from the transfer function $H$ but from the forcing $\widehat{V}(s)$, and corresponding to the steady-state solution. Indeed, computing the residue with respect to that pole gives the equilibrium distribution:
	\begin{align}
		\label{eq:rhoeq}
		\rho_{eq}(z) &=
		\text{Res}_{\widehat{\rho}(z,s)e^{st}}(s=0)
		=\lim_{s\rightarrow 0}(s\widehat{\rho}(z,s)e^{st})
		\nonumber \\ 
		&= -v\frac{\sinh\frac{z}{\epsilon}}{\sinh\frac{1}{\epsilon}} \; ,
	\end{align}
	where we have used Eqs.~\ref{eq:rhoLaplace} and~\ref{eq:defH}.
	The full transient response from Eq.~\ref{eq:rhoresidues} can then be rewritten, using Eq.~\ref{eq:rhoeq}, as
	\begin{align}
		\rho(z,t)
		& = \rho_{eq}(z) + \sum_{s\in\{s_n^{\Delta V}\}}
		\text{Res}_{\widehat{\rho}(z,s)e^{st}}(s)
		\nonumber\\
		&\hspace{0.1cm}
		+ 2\sum_{n=0}^\infty \frac{s_n(1+\epsilon s_n)}{3-s_n^2}\frac{\sinh\frac{z\sqrt{1+\epsilon s_n}}{\epsilon}}{\sinh\frac{\sqrt{1+\epsilon s_n}}{\epsilon}}\widehat{V}(s_n)e^{s_nt}
		\label{eq:relaxation_anyV}
	\end{align}
	where the first sum runs over poles of $\widehat{\Delta V}$ and the second over that of the transfer function $H$ (for the derivation of the last term, see Refs.~\cite{Janssen2018, palaia_pnp_2025}). 
	
	Similarly, the surface charge density can be expressed in the time domain, using Eq.~\ref{eq:sigmaLaplace}, as
	\begin{align}
		\sigma(t)&=\sigma_{eq} + \sum_{s\in\{s_n^{\Delta V}\}}
		\text{Res}_{\widehat{\sigma}(s)e^{st}}(s)
		\nonumber\\
		&+ \frac{2}{\epsilon}\sum_{n=0}^\infty\frac{1+\epsilon s_n}{s_n^2-3}\widehat{V}(s_n)e^{s_nt},
		\label{eq:sigma_anyV}
	\end{align}
	where the equilibrium value  
	\begin{align}
		\sigma_{eq} &= -\frac{v}{\epsilon}\coth\frac{1}{\epsilon}
		\label{eq:sigmaeq}
	\end{align}
	allows one to define the capacitance $\sigma_{eq}/v$ of the system.
	
	The ionic density profiles can be deduced from that of the charge in the linear regime (where in reduced units $n_++n_-=2$ and $n_+-n_-=2\rho$) as:
	\begin{equation}
		\label{eq:ionic_in_chargedensity}
		n_\pm(z,t)=1\pm\rho(z,t) \; .
	\end{equation}
	This allows us in particular to quantify the deviation of the system from its equilibrium state reached for $t\to\infty$. Inspired by the notion of Kullback–Leibler (KL) divergence~\cite{kullback_information_1951}, we normalize the density profiles (by $2n_0L$ so that their integrals are equal to 1) to introduce
	\begin{align}
		D_{KL}(t) &=\frac{1}{2n_0L}\int_{-L}^L\left[n_{+}^{eq}(z)\ln\frac{n_{+}^{eq}(z)}{n_+(z,t)} \right. \nonumber \\
		& \left. \hspace{2cm} +n_{-}^{eq}(z)\ln\frac{n_{-}^{eq}(z)}{n_-(z,t)}\right] \, {\rm d}z \nonumber \\
		&=\frac{1}{2n_0L}\int_{-L}^L\left[(1+\rho_{eq}(z))\ln\frac{1+\rho_{eq}(z)}{1+\rho(z,t)} \right. \nonumber \\
		& \left. \hspace{1cm} +(1-\rho_{eq}(z))\ln\frac{1-\rho_{eq}(z)}{1-\rho(z,t)}\right] \, {\rm d}z \, .
		\label{eq:KL_ionic}
	\end{align}
	This object allows to quantity the completeness of convergence towards the steady state distributions for both cation and anion densities. When the profiles have reached equilibrium, $D_{KL}$ vanishes. Note that by construction, such a quantity is always positive.

	\subsection{Shortcutting the charging time}
	\label{sec:response:shortcut}
	
	Given that $\widehat{\rho}(z,s) \propto H(s)\widehat{V}(s)$, we can use $V$ to eliminate poles of $H$ that correspond to slow relaxation modes. This can be done by simply prescribing that $V$ has a zero where $H$ has a pole. The solution Eq.~\ref{eq:relaxation_anyV} for an arbitrary driving potential of the form Eq.~\ref{eq:defDeltaV} thus provides a way to relax to the equilibrium density profile (and corresponding equilibrium charge of the electrodes) faster than with a step potential, which corresponds to a driving $\Delta V(t)=0$ and a solution for the charge density profile with $\widehat{V}(s_n)=v/s_n$ in Eq.~\ref{eq:relaxation_anyV}. By choosing a driving $V(t)$ that does not introduce additional poles (other than $s=0$ corresponding to the equilibrium) and such that $\widehat{V}(s_n)=0$ for $n\in\{0,...,m-1\}$, we ensure that the $m$ slowest modes of the density do not contribute to the relaxation dynamics. Indeed, for such a choice Eq. \eqref{eq:sigma_anyV} reduces to
	\begin{align}
		\sigma(t) &= \sigma_{eq} 
		+ \frac{2}{\epsilon}\sum_{n=m}^\infty\frac{1+\epsilon s_n}{s_n^2-3}\widehat{V}(s_n)e^{s_nt}
		\label{eq:surfacechargedensity_shortcut_evolution}
	\end{align}
	and, similarly, for the density profiles Eq.~\ref{eq:relaxation_anyV} reduces to
	\begin{align}
		\rho(z,t)&= \rho_{eq}(z)
		\nonumber\\
		&\hspace{-0.2cm}
		+2\sum_{n=m}^\infty \frac{s_n(1+\epsilon s_n)}{3-s_n^2}\frac{\sinh\frac{z\sqrt{1+\epsilon s_n}}{\epsilon}}{\sinh\frac{\sqrt{1+\epsilon s_n}}{\epsilon}}\widehat{V}(s_n)e^{s_nt}
		\; .
		\label{eq:chargedensity_shortcut_evolution}
	\end{align}
	
	In order to find a suitable protocol satisfying the cancellation of $m$ modes ($\widehat{V}(s_n)=0$ for $n\in\{0,...,m-1\}$) as well as the initial and final conditions ($V(0)=0$, $V(t_f)=v$), \textit{i.e.} a total of $m+2$ constraints, we look for a combination of the form $\Delta V(t)=v\sum_{i=0}^{m+1} a_i f_i(t)$,  where $f_i(t)$ are $m+2$ linearly independent functions such that they do not all vanish at $t=0$ and that they all vanish for $t>t_f$. We further introduce the Laplace transforms (see Eq.~\ref{eq:defLaplace}, recalling that $f_i$ vanishes for $t>t_f$):
	\begin{equation}
		\label{eq:elli}
		\ell_i(s,t_f)=\int_0^{t_f} f_i(t)e^{-st}dt \; .
	\end{equation}
	The mentioned constraints can then be expressed as a system of linear equations for the coefficients $a_i$:
	\begin{equation}
		\begin{bmatrix}
			f_0(0)&...&f_{m+1}(0)\\
			f_0(t_f)&...&f_{m+1}(t_f)\\
			\ell_0(s_0,t_f)&...&\ell_{m+1}(s_0,t_f)\\
			:&:&:\\
			\ell_0(s_{m-1},t_f)&...&\ell_{m+1}(s_{m-1},t_f)
		\end{bmatrix}
		\begin{bmatrix}
			a_0\\a_1\\a_2\\:\\a_{m+1}
		\end{bmatrix}
		=
		\begin{bmatrix}
			-1\\0\\-\frac{1}{s_0}\\:\\-\frac{1}{s_{m-1}}
		\end{bmatrix}
		\label{eq:matrixequation}
	\end{equation}
	where the first two rows impose the initial and final conditions, and the remaining rows ensure pole cancellation. For a given choice of suitable functions $f_i(t)$, the $m+2$ coefficients $a_i$ obtained by inverting the matrix in Eq.~\ref{eq:matrixequation} define a protocol allowing us to eliminate the $m$ slowest modes from the relaxation towards the equilibrium, which are all present when a step function is used.

	\subsection{Polynomial driving}
	\label{sec:response:polynomialdriving}
	
	We now illustrate the proposed approach to shortcut the charging dynamics, using the specific choice of polynomial driving for $\Delta V(t)$, \textit{i.e.} $f_i(t)=t^i$ for $t\in[0,t_f]$ and 0 elsewhere. The Laplace transform
	\begin{equation}
		\label{eq:vhatpolynomial}
		\widehat{V}(s)= \frac{v}{s} + v\sum_{i=0}^{m+1}\dfrac{a_i}{s^{i+1}}\left[i!-\Gamma(1+i,st_f) \right] \;,
	\end{equation}
	where $\Gamma$ is the upper incomplete Gamma function, reveals no pole apart from $s=0$, so that with the proper choice of coefficients the charge density will evolve according to Eq.~\ref{eq:chargedensity_shortcut_evolution}. For polynomial driving, the boundary condition $V(0)=0$ leads to $a_0=-1$, and the other coefficients can be obtained by solving Eq.~\ref{eq:matrixequation} numerically. In practice, the infinite sum over modes $s_n$ in Eq.~\ref{eq:chargedensity_shortcut_evolution} has to be truncated at a finite $n_{\rm max}\gg m$ value, which can be chosen to achieve a prescribed numerical precision thanks to the fast decay with $n$. Results shown below were obtained by truncation at $n_{\rm max}=10^2$.

	\subsubsection{Driving potential and surface charge density}
	\label{sec:response:polynomialdriving:potentialsurfacecharge}
	
	\begin{figure}[ht!]
		\centering
		\includegraphics[width=.9\columnwidth]{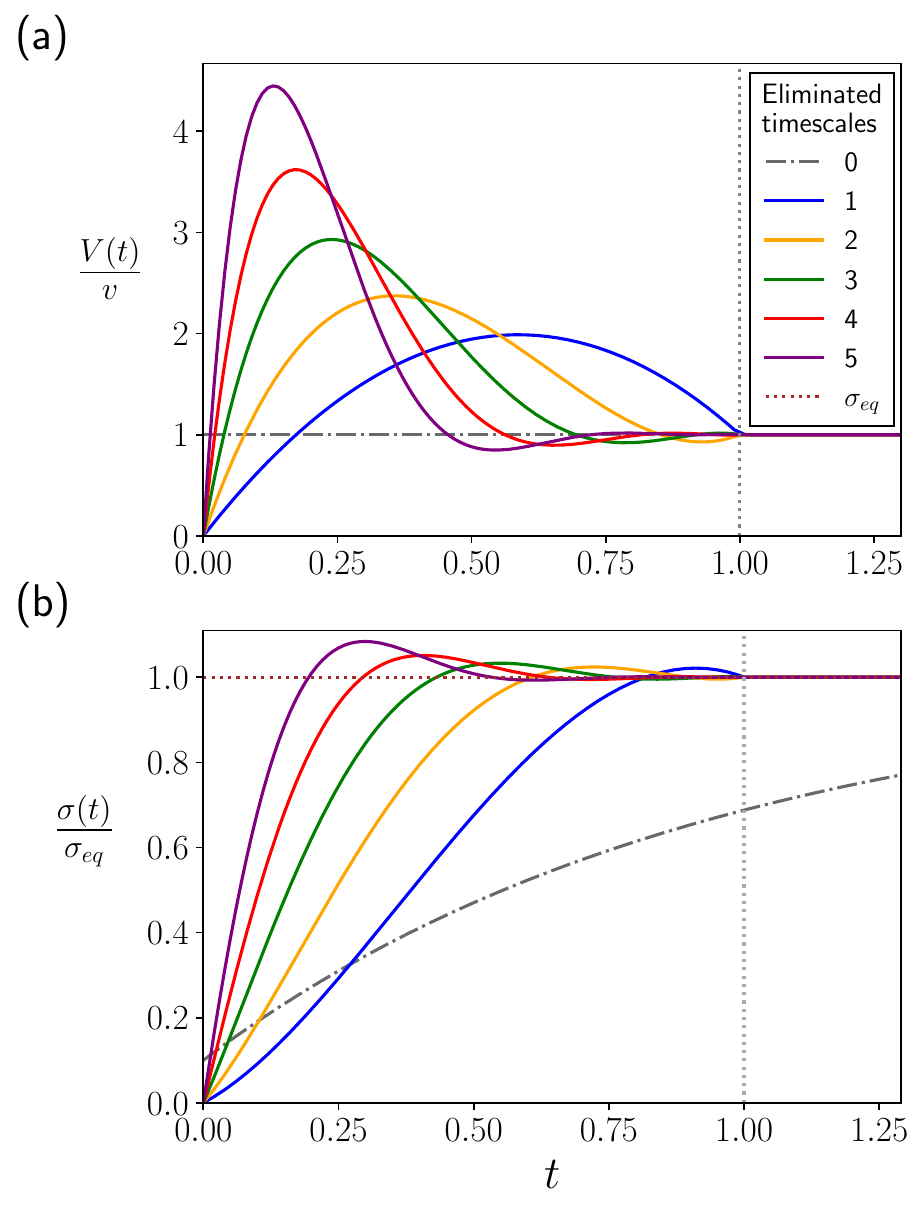}
		\caption{(a) Polynomial driving $V(t)$, normalized by the reduced voltage at steady-state $v$, for a reduced Debye length $\epsilon=0.1$ and reduced driving time $t_f=1$ (see Table~\ref{tab:units}), with coefficients obtained by solving Eq.~\ref{eq:matrixequation}, for $m=1$ to 5  timescales eliminated. (b) Evolution of the surface charge density, $\sigma(t)$, normalized by its equilibrium value (see Eq.~\ref{eq:sigmaeq}) to the polynomial drivings; the dashed-dotted lines indicate the response to the voltage step $v\,\Theta(t)$. In all panels, the vertical dotted gray lines at $t=t_f=1$ mark the end of the driving. The situation with 0 eliminated time scales corresponds to driving with a step potential (slowest response in $\sigma$).
		}
		\label{fig:polynomial_driving_sigma}
	\end{figure}

	Fig.~\ref{fig:polynomial_driving_sigma}a shows the polynomial driving potentials obtained by the procedure described above to eliminate 1 to 5 relaxation modes, for a reduced Debye length $\epsilon=0.1$, a reduced voltage at steady-state $v$ and a reduced driving time $t_f=1$. This corresponds to relatively thin electric double layers, a low potential as requested to stay in the linear response regime, and a final driving time comparable to the $RC$ charging time. As explained in the previous section, eliminating $m$ modes requires a polynomial of order $m+1$. Qualitatively, accelerating the relaxation requires larger voltages than the steady-state value, as could have been anticipated. Increasing the order of the polynomial results in a drive with an increasing initial overshoot and a faster return to the steady-state voltage, via (slightly) lower voltages, with a smaller slope while approaching the final driving time. This will be discussed in more detail in Section~\ref{sec:response:polynomialdriving:drivingtime} below.
	
	The corresponding evolution of the surface charge density $\sigma(t)$ is shown in Fig.~\ref{fig:polynomial_driving_sigma}b. In all cases, the proposed driving allows one to reach the target equilibrium value $\sigma_{eq}$ at the end of the protocol ($t=t_f=1$), whereas this is not the case for the step voltage (dashed-dotted line, corresponding to 0 eliminated time scales). The overshoot in driving potential visible in Fig.~\ref{fig:polynomial_driving_sigma}a results in a less pronounced overshoot of the surface charge. As expected, eliminating an increasing number of relaxation timescales (via a stronger initial driving) results in a faster convergence to equilibrium. We note that while we imposed $V(t=0)=0$, resulting in a vanishing surface charge, after switching the step potential the surface charge jumps to a finite value (see the beginning of the dashed-dotted line in Fig.~\ref{fig:polynomial_driving_sigma}b) to a uniform electric field throughout the uncharged liquid, as the ions have not yet moved~\cite{Asta2019JCP}.

	\subsubsection{Charge and ionic density profiles}
	\label{sec:response:polynomialdriving:chargeions}
	
	\begin{figure}[ht!]
		\centering
		\includegraphics[width=.9\columnwidth]{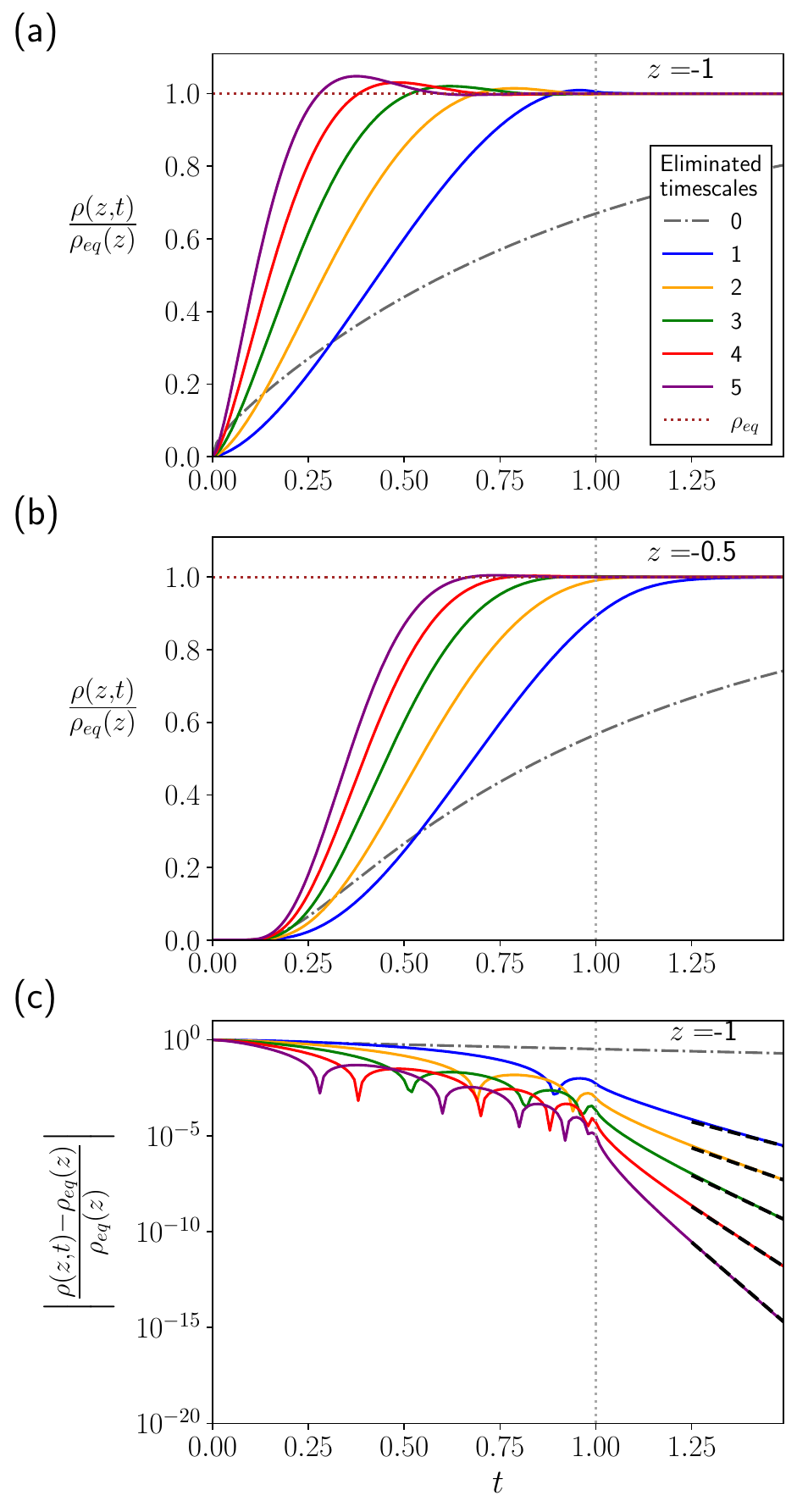}
		\caption{(a) and (b) Evolution of the charge density, $\rho(z,t)$ to the polynomial drivings at reduced positions $z=-1$ (at the left electrode) and $z=-0.5$, respectively, for the same conditions as in Fig.~\ref{fig:polynomial_driving_sigma}, normalized by the equilibrium densities $\rho_{eq}(z)$.
			(c) Relative deviation from the equilibrium charge density at the left electrode surface ($z=-1)$, plotted on a logarithmic scale, highlighting in particular the exponential decay at long times; the slopes of the black dashed lines in this regime are the smallest remaining poles $s_m$ of the transfer function (see Eq.~\ref{eq:defH}) after eliminating $m$ modes (see Eq.~\ref{eq:chargedensity_shortcut_evolution}).
			In all panels, the dashed-dotted lines indicate the response to the voltage step $v\,\Theta(t)$ and  the vertical dotted gray lines at $t=t_f=1$ mark the end of the driving.
		}
		\label{fig:polynomial_driving_rho}
	\end{figure}
	
	Fig.~\ref{fig:polynomial_driving_rho}a and~\ref{fig:polynomial_driving_rho}b show the evolution of the charge-density, $\rho(z,t)$ to these polynomial drivings at reduced positions $z=-1$ (\textit{i.e.} at the left electrode) and $z=-0.5$, respectively. The response to the reference voltage step $v\,\Theta(t)$ is also indicated as dashed-dotted lines, and the equilibrium densities $\rho_{eq}(z)$ are shown as horizontal dotted lines. Both panels demonstrate that, while the local charge density is still far from its equilibrium value with 0 eliminated timescales at $t=t_f=1$ (vertical dotted lines), the proposed polynomial drivings result in densities much closer to the equilibrium value, and that this plateau is reached faster with increasing number of eliminated time scales, as expected. 
	
	We further note that equilibrium is reached faster close to the surface ($z=-1$) than away from it ($z=-0.5$). This is expected, since the system is driven at its boundaries, and the field is progressively screened by the build-up of the electric double layers, leading to weaker fields acting on the ions beyond it. As a result, even though the same relaxation modes $s_n$ contribute to the evolution at all positions $z$, their amplitudes depend on the position (as can be seen from Eq.~\ref{eq:chargedensity_shortcut_evolution}). In particular, Fig.~\ref{fig:polynomial_driving_rho}a clearly shows that the charge density near the wall does not vary monotonically with time, due to the initial potential overshoot. While this may also be the case further from the surface, this is not obvious from Fig.~\ref{fig:polynomial_driving_rho}b. 
	
	In order to further characterize this phenomenon, Fig.~\ref{fig:polynomial_driving_rho}c shows the relative deviation from the equilibrium charge density at the left electrode surface ($z=-1)$. The logarithmic scale first shows that the proposed protocols allow one to reach a charge density at the final driving time $t_f$ that is orders of magnitude closer to the equilibrium value than with the voltage step (dashed-dotted line). In addition, the cusps before $t_f$, whose number is equal to that of eliminated time scales, reveal the oscillations around the equilibrium value during the driving. Finally, the logarithmic scale also highlights the exponential decay at long times (dashed black lines), with a rate equal to the smallest remaining pole $s_m$ of the transfer function (see Eq.~\ref{eq:defH}) after eliminating $m$ modes (see Eq.~\ref{eq:chargedensity_shortcut_evolution}).
	
	\begin{figure}[ht!]
		\centering
		\includegraphics[width=0.9\columnwidth]{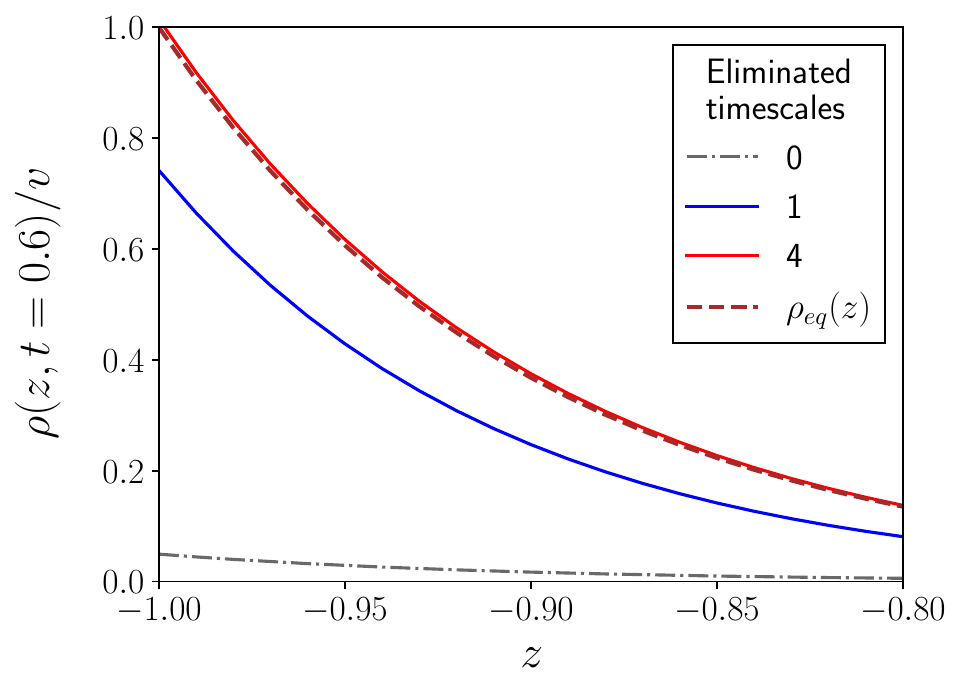}
		\caption{Charge-density profile $\rho(z,t)$ near the left electrode, divided by the steady-state voltage $v$, at reduced time $t=0.6$ under the action of various drivings, for $\epsilon=0.1$ and $t_f=1$: for a potential step (dashed-dotted line) and for polynomial drivings with $m=1$ to 4 relaxation modes eliminated (full lines). The dashed line indicates the equilibrium density profile.
		}
		\label{fig:chargedensity_whole_profile}
	\end{figure}
	
	We now turn to the global response of the system. Fig.~\ref{fig:chargedensity_whole_profile} shows the charge-density profile $\rho(z,t)$ near the left electrode, divided by the steady-state voltage $v$, at a fixed reduced time $t=0.6$ under the action of various drivings, for $\epsilon=0.1$ and $t_f=1$. For a potential step (dashed-dotted line), the profile is still far from the equilibrium charge distribution (dashed line). With one mode eliminated, the profiles get closer, and with four modes eliminated, the profiles cannot be distinguished from equilibrium on this scale. These results clearly demonstrate that the proposed drivings accelerate the build-up of the electric double layer at the electrode-electrolyte interface.

	\begin{figure}[ht!]
		\centering
		\includegraphics[width=0.95\columnwidth]{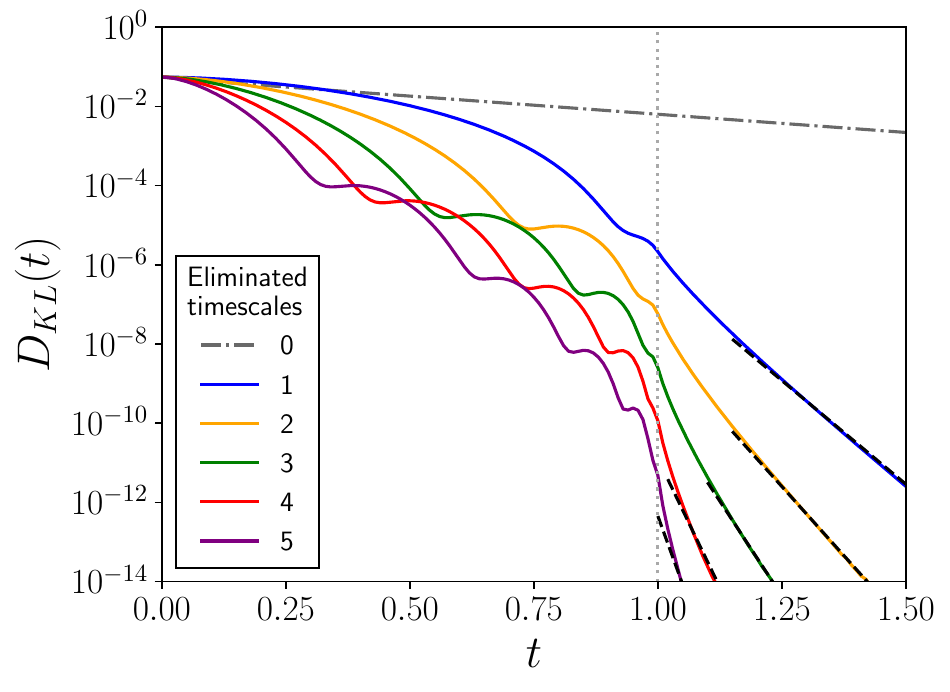}
		\caption{
			Global measure $D_{KL}(t)$ (see Eq.~\ref{eq:KL_ionic}) of the deviation of the ionic density profiles from their equilibrium distribution as a function of time, for $\epsilon=0.1$, $v=0.1$ and $t_f=1$, in response to a voltage step (dashed-dotted lines) and polynomial drivings designed to eliminate $m=1$ to 5  relaxation modes (solid lines). The vertical dotted lines mark the end of the driving at $t=t_f=1$. The logarithmic scale highlights in particular the exponential decay at long times and the slopes of the black dashed lines in this regime are twice the smallest remaining poles $s_m$ of the transfer function (see Eq.~\ref{eq:defH}) after eliminating $m$ modes (see Eq.~\ref{eq:logKL_ionic}). 
		}
		\label{fig:KLdiv_ionic}
	\end{figure}
	
	Finally, we consider the global measure $D_{KL}(t)$, introduced in Eq.~\ref{eq:KL_ionic}, of the deviation of the ionic density profiles from their equilibrium distribution, shown in Fig.~\ref{fig:KLdiv_ionic}.  This global measure displays features similar to those of Fig.~\ref{fig:polynomial_driving_rho}c for a single position. Importantly, $D_{KL}(t)$ at the final driving time $t_f$ is orders of magnitude closer to the equilibrium value than with the voltage step, with a decrease by 1 or 2 orders of magnitude for each additional relaxation time eliminated. The logarithmic scale also highlights the exponential decay at long times, with a rate equal to twice the smallest remaining poles $s_m$ of the transfer function (see Eq.~\ref{eq:defH}) after eliminating $m$ modes (see Eq.~\ref{eq:chargedensity_shortcut_evolution}). Indeed, assuming that at long times $\delta\rho(z,t)=\rho(z,t)-\rho_{eq}(z)$ is small and dominated by the slowest remaining mode, one can expand Eq.~\ref{eq:KL_ionic} in $\delta\rho$ to show that
	\begin{equation}
		\label{eq:logKL_ionic}
		\ln D_{KL}(t) \approx 2s_m t + c_m \; ,
	\end{equation}
	with $c_m$ a constant that depends on the number of eliminated modes. The factor of 2 arises from the fact that, in this limit, the leading order of $D_{KL}(t)$ is quadratic in $\delta\rho$ and this scaling is confirmed in Fig.~\ref{fig:KLdiv_ionic}. This emphasizes that our protocol is globally, rather than just locally, efficient and that it speeds up relaxation throughout the whole electrolyte.

	\subsubsection{Effect of the driving time}
	\label{sec:response:polynomialdriving:drivingtime}
	
	\begin{figure}[ht!]
		\centering
		\includegraphics[width=0.95\columnwidth]{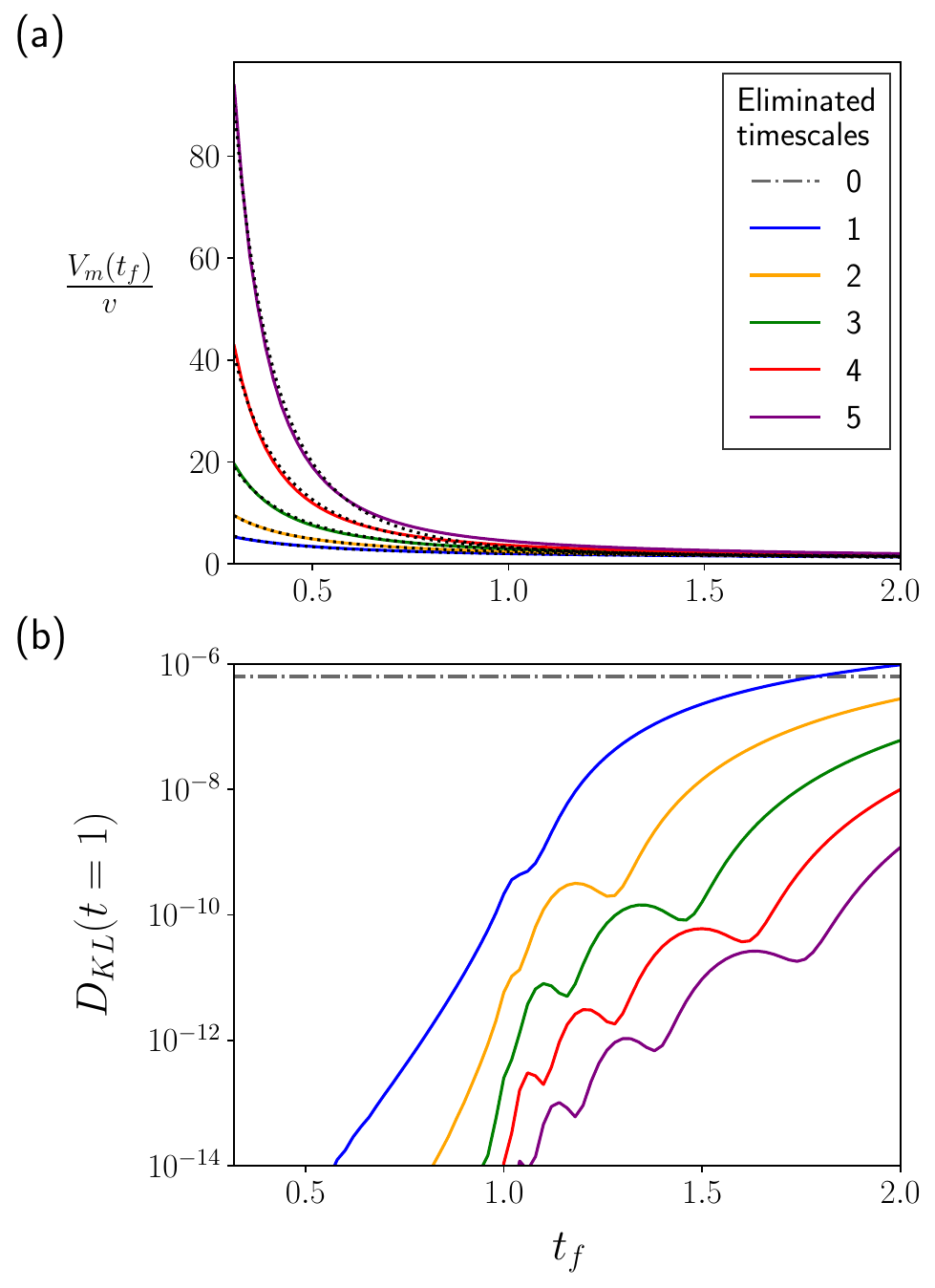}
		\caption{(a) Maximum voltage reached during the driving process for polynomial drivings designed to eliminate $m=1$ to 5  relaxation modes (see Fig.~\ref{fig:polynomial_driving_sigma}a), as a function of variation of the reduced driving time $t_f$, for fixed $\epsilon=0.1$ and $v=10^{-3}$. Dotted lines indicate fits of the form $V_m/v = 1 + c/t_f^\alpha$. (b)  Global measure $D_{KL}(t=1)$ (see Eq.~\ref{eq:KL_ionic}) of the deviation of the ionic density profiles from their equilibrium distribution at a fixed reduced time $t=1$, for drivings eliminating 1 to 5 relaxation modes. The horizontal dashed-dotted line indicates the result for the voltage step.}
		\label{fig:drivingtime}
	\end{figure}
	
	All the above results correspond to a reduced driving time $t_f=1$ in units of $L\lambda_D/D$, the $RC$ time in the thin EDL limit, approximately equal to the slowest relaxation mode $-1/s_0$ for a reduced Debye length $\epsilon=0.1$ (see Section~\ref{sec:response:relaxationdynamics}). They show that it is possible to achieve faster equilibration with driving protocols with transient larger voltages, with a maximum, $V_m$, increasing with the number of relaxation modes eliminated (see Fig.~\ref{fig:polynomial_driving_sigma}a). One might also expect that a faster charging can be obtained by decreasing the driving time $t_f$, by construction. Fig.~\ref{fig:drivingtime}a shows that this also requires voltages that increase rapidly with decreasing $t_f$ (the results were obtained for $v=10^{-3}$ to ensure that we remain in the linear response regime). The maximum seems to approximately diverge as $V_m/v \approx 1+c/t_f^\alpha$ (dotted lines) in the considered range of $t_f$. We find numerically that the constant $c$ varies slightly between 1 and 2 with the number $m$ of eliminated timescales, while the exponent grows approximately as $\alpha\approx(m+1)/2$ for $m>1$. This highlights the rapid increase in the maximum voltage with both increasing $m$ and decreasing $t_f$.  
	
	Finally, Fig.~\ref{fig:drivingtime}b shows the global measure $D_{KL}(t=1)$ (see Eq.~\ref{eq:KL_ionic}) of the deviation of the ionic density profiles from their equilibrium distribution at a fixed reduced time $t=1$, for drivings eliminating 1 to 5 relaxation modes. The results indicate that several combinations of driving times $t_f$ and number of eliminated modes can achieve the same deviation from equilibrium at a fixed target time $t=1$. In fact, one can reach equilibrium faster than under a voltage step even with driving times $t_f$ exceeding the $RC$ time, by eliminating a sufficient number of relaxation modes. For example, for $t_f=2$ the response to a step voltage (gray dashed-dotted line) is closer to equilibrium at $t=1$ than if one eliminates only one timescale (blue line), but two (yellow line) or more timescales bring the system closer to equilibrium at that time. This offers some flexibility to shortcut the dynamics if one wishes to introduce additional constraints or objectives, such as not exceeding a threshold for the maximum voltage $V_m$ or minimizing the energy dissipated during the charging process.

	\subsubsection{Thin EDL limit}
	\label{sec:response:polynomialdriving:thinEDL}
	
\begin{figure}[ht!]
		\centering
		\includegraphics[width=0.95\columnwidth]{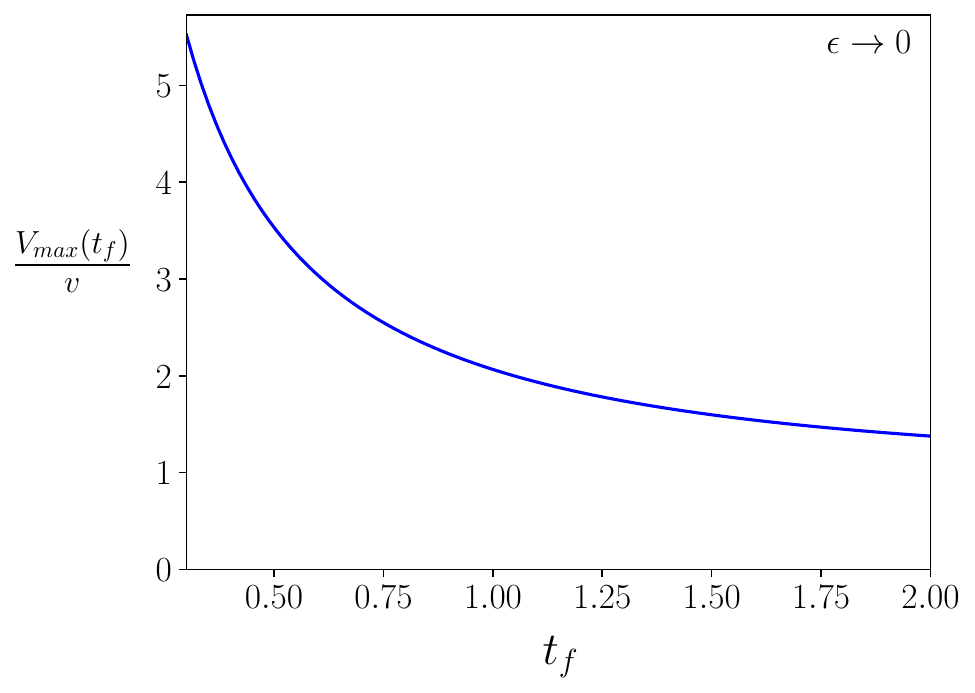}
		\caption{Maximum voltage reached with polynomial driving to eliminate the slowest mode ($m=1$) in the thin EDL ($\epsilon\to0$) limit, $V_{\text{max}}$, given as a function of the reduced driving time $t_f$ by Eq.~\ref{eq:peakpotential1elimination_thinEDL} with $s_0=-1$.}
		\label{fig:potentialpeak1elimination}
	\end{figure}

	In the thin EDL limit ($\epsilon\rightarrow0$), $s_0\to-1$, corresponding to a relaxation time $L\lambda_D/D$ and all other modes collapse to $s_1\to-1/\epsilon$, corresponding to a relaxation time $\lambda_D^2/D$ \cite{palaia_charging_2025,palaia_pnp_2025,Palaia2019}. The collapse of all modes $n\geq1$ suggests that significant speed-up could be achieved by eliminating a single mode $s_0$, since the remaining one would be faster by a factor of $1/\epsilon$. This can be achieved by a polynomial driving (see Section~\ref{sec:response:polynomialdriving}) by the parabolic form until $t_f$
	\begin{equation}
		\frac{V(t)}{v} 
		=\frac{t}{t_f}\frac{(e^{s_0t_f}-1)s_0t + 2(1+s_0t_f-e^{s_0t_f})}{2+s_0t_f+(s_0t_f-2)e^{s_0t_f}}
		\label{eq:rescaleddriving1elimination}
	\end{equation}
	that reaches a maximum at
	\begin{equation}
		t_{\text{max}}=\frac{t_f}{1-e^{s_0t_f}}+\frac{1}{s_0}
		\; ,
		\label{eq:tmax1elimination}
	\end{equation}
	with a value
	\begin{equation}
		\frac{V(t_{\text{max}})}{v}
		=\frac{(1+s_0t_f-e^{s_0t_f})^2}{s_0t_f(e^{s_0t_f}-1)(2+s_0t_f+(s_0t_f-2)e^{s_0t_f})}
		\label{eq:peakpotential1elimination_thinEDL}
	\end{equation}
	where, as mentioned above, $s_0\to-1$ in this limit. The maximum occurs at a time (see Eq.~\ref{eq:tmax1elimination}) between $t_f/2$ and $t_f$, and the maximum value required to eliminate the slowest mode decreases with increasing driving time $t_f$, as shown in Fig.~\ref{fig:potentialpeak1elimination}. 
	
	In the thin EDL limit ($\epsilon\to0$), the response to a voltage step is to leading order
	\begin{equation}
		\frac{\sigma(t)}{\sigma_{eq}}\approx 1-e^{-t}
		\; ,
		\label{eq:sigmaepsilon0steprelaxation}
	\end{equation}
	\textit{i.e.} a mono-exponential relaxation with a characteristic time $L\lambda_D/D$, as expected. The response to the polynomial driving Eq.~\ref{eq:rescaleddriving1elimination} eliminating this timescale can be determined asymptotically from Eq.~\ref{eq:surfacechargedensity_shortcut_evolution}. To that end, one can write the poles for $n\geq1$ as $s_n=-\frac{1}{\epsilon}+\epsilon u_n$ with $-\left(n+\frac{1}{2}\right)^2\pi^2<u_n<-n^2\pi^2$ (see Section~1.4.1 of Ref.~\citenum{Palaia2019}), and then approximate the series using $u_n\approx -n^2\pi^2$ and taking the limit $\epsilon\to0$. This leads, for $t>t_f$ to
	\begin{equation}
		\frac{\sigma(t)}{\sigma_{eq}}
		\approx 1 + g(t_f) \frac{\epsilon^{9/2} \, e^{-(t-t_f)/\epsilon}}{(t-t_f)^\frac{3}{2}}
		\label{eq:sigmaepsilon0shortcut}
	\end{equation}
	with 
	\begin{equation}
		g(t_f) = \frac{1}{2\sqrt{\pi}} \left( \frac{1}{t_f\coth\frac{t_f}{2}-2}-\frac{1}{t_f}  \right) 
		\; .
		\label{eq:sigmaepsilon0shortcutprefactor}
	\end{equation}
	The non-exponential dependence arises from the collapse of modes $n\geq1$ in the $\epsilon\to0$ limit. Numerically, Eq.~\ref{eq:sigmaepsilon0shortcut} describes very well the results for times $t>t_f$ in the case $\epsilon=10^{-2}$ (not shown).

	\subsubsection{Beyond the linear response regime}
	\label{sec:response:polynomialdriving:beyondlinear}
	
	The present strategy to eliminate relaxation modes relies on the analysis of the linear response of the system. Nevertheless, one may anticipate that the protocols derived in the $v\ll1$ limit could accelerate the charging process also for moderate voltages, albeit less efficiently. To illustrate this, we solve numerically the PNP equations to predict the response of the system for a reduced Debye length $\epsilon=0.1$ to the polynomial driving obtained by solving Eq.~\ref{eq:matrixequation} to eliminate $m=1$ timescale within a driving time $t_f=1$ in the linear response regime, for larger reduced voltages $v=1$ and $v=2$.
	
	\begin{figure}[ht!]
		\centering
		\includegraphics[width=0.95\columnwidth]{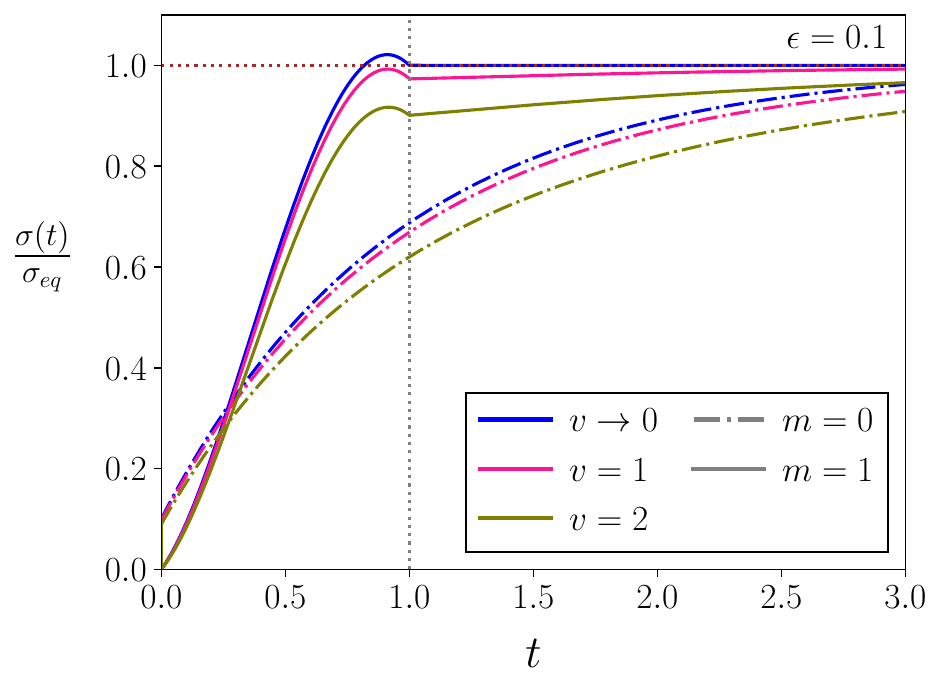}
		\caption{
			Evolution of the surface charge density, $\sigma(t)$, normalized by its equilibrium value (see Eq.~\ref{eq:sigmaeq}), in response to the polynomial driving for a reduced Debye length $\epsilon=0.1$ and reduced driving time $t_f=1$ (see Table~\ref{tab:units}), with coefficients obtained by solving Eq.~\ref{eq:matrixequation} to eliminate $m=1$ timescale in the linear response regime (solid lines); the dashed-dotted lines indicate the response to the voltage step $v\,\Theta(t)$, corresponding to $m=0$ timescale eliminated. Results are shown for the linear response limit $v\to0$ (blue lines), $v=1$ (pink lines) and $v=2$ (olive lines). While the same polynomial driving is applied in the three cases, the response is predicted by Eq.~\ref{eq:surfacechargedensity_shortcut_evolution} for $v\to0$ and by solving numerically the PNP equations for $v=1$ and $v=2$.
			The vertical dotted gray line at $t=t_f=1$ marks the end of the driving.
		}
		\label{fig:nonlinearesponse}
	\end{figure}
	
	The results, shown in Fig.~\ref{fig:nonlinearesponse}, demonstrate that the same protocol ($m=1$, solid lines) achieves a faster charging than the step voltage ($m=0$, dashed-dotted lines), even though not as efficiently as in the linear response limit. After the driving time $t_f=1$, the charge evolves toward the equilibrium value following an approximately exponential form with a voltage-dependent rate. For a voltage step, the $RC$ time is expected to increase in the present moderately non-linear regime as $\propto \cosh(v/2)$ (see Refs.~\citenum{palaia_charging_2025, palaia_pnp_2025}). The slower decay to the plateau observed for increasing $v$ is consistent with this prediction. Despite this slower decay beyond $t_f$, the value at $t_f$ is already closer to the plateau value than with the step voltage. This illustrates that the present strategy derived in the linear response regime may prove useful for moderate voltages. Nevertheless, other approaches should be considered for a more efficient speed-up in this case.

	\section{Conclusion}
	
	In this work, we investigated the possibility of accelerating the charging dynamics of electric double-layer capacitors (EDLCs) using time-dependent voltage protocols, inspired by the concept of “shortcuts to adiabaticity” (even though the present approach differs from the standard methods in this context). Within the Poisson-Nernst-Planck model and the linear response regime, we derived analytical protocols capable of eliminating an arbitrary number of relaxation modes, thereby enabling the system to approach its equilibrium charged state within a finite time ---potentially orders of magnitude faster than the intrinsic relaxation time (provided that the system remains in the linear regime). Our approach relies on designing driving protocols that suppress the slowest relaxation modes, allowing for rapid and controlled charging. Through this method, we demonstrated that the surface charge density, charge-density profiles, and global deviation from equilibrium (quantified by a Kullback–Leibler-like divergence) can all be significantly accelerated, even for driving times comparable to or shorter than the natural $RC$ time of the system.
	
	The results highlight the efficacy of these protocols in achieving near-equilibrium states at the electrode-electrolyte interface, both locally and globally. The elimination of relaxation modes not only reduces the time required to reach equilibrium but also provides flexibility in optimizing the charging process under additional constraints, such as limiting the maximum applied voltage or minimizing energy dissipation. The present approach could in principle be extended to take into account many non-ideal effects that could be present in the system, such as electrostatic correlations \cite{electrostaticcorr10.1063/5.0068521, PhysRevE.86.056303, doi:10.1021/acs.jctc.2c00607, D0SM01523G, doi:10.1021/acs.jpcb.2c00028, palaia_pnp_2025}, chemical reactions \cite{doi:10.1021/acs.jpcc.8b10473, doi:10.1021/acs.jpca.9b00302}, solvation effects \cite{solvationeffects}, or advection in the electrolyte. While the present study focuses on planar EDLCs and symmetric electrolytes, the methodology and insights offer promising avenues for enhancing the performance of energy storage devices and other applications where rapid charging dynamics are critical, such as digital memory systems. In the context of supercapacitors, the present strategy could also be followed by including the transport of ions inside porous electrodes at a comparable level of description, \textit{e.g.} as in Ref.~\citenum{Lian2020}, even though more elaborate models might be required to capture dynamical effects due to crowding or solvation in nanopores observed in molecular simulations~\cite{pean_dynamics_2014, pean_confinement_2015, pean_multi-scale_2016, Breitsprecher2018, Breitsprecher2020}. This work underscores the potential of time-dependent voltage protocols as a powerful tool for controlling the dynamics of electrochemical systems by systematically eliminating slow modes.
	
	\section*{Acknowledgments}
	
	This work has received funding from the European Union's Horizon 2020 and Horizon Europe research and innovation programs under the Marie Sk\l{}odowska-Curie grant agreement 101119598-FLUXIONIC, as well as from the European Research Council (grant agreement no.~863473).

	\section*{Author declarations}
	
	\subsection*{Conflict of interest}
	There is no conflict of interest to declare.
	
	\subsection*{Author contributions}
	\textbf{Megh Dutta} Conceptualization (equal); Formal analysis (equal); Investigation (lead); Methodology (equal); Writing/Original Draft Preparation (lead); Writing – review \& editing (equal). \textbf{Ivan Palaia} Conceptualization (equal); Formal analysis (equal); Investigation (supporting); Methodology (equal); Writing – review \& editing (supporting). \textbf{Emmanuel Trizac:} Conceptualization (equal); Formal analysis (equal); Funding Acquisition (equal); Investigation (supporting); Methodology (equal); Supervision (equal); Writing/Original Draft Preparation (supporting); Writing – review \& editing (lead). \textbf{Benjamin Rotenberg:} Conceptualization (equal); Formal analysis (equal); Funding Acquisition (equal); Investigation (supporting); Methodology (equal); Supervision (equal); Writing/Original Draft Preparation (supporting); Writing – review \& editing (lead).

	\section*{Data availability}
	The original data presented in this study are openly available in Zenodo at 
	\href{https://zenodo.org/records/21824496?token=eyJhbGciOiJIUzUxMiJ9.eyJpZCI6IjMwZjc1MjNjLTNjMDktNDE1Yi04ZDg5LTBhMTBjYTYyYjk5ZSIsImRhdGEiOnt9LCJyYW5kb20iOiIxYTFiODM5ZTk3OTEyYWU2YmM2OGU3OWZhMDcwYmVhOSJ9.RlEj4rw40-xx-y7c_Tp6NSwkkVv2UGEGCXT2I-FpunW-Vq4bjrKIkQW3DITCwz4uNPd8wfAEK4RIKG8gAhfOSA}{10.5281/zenodo.21824495}.

\end{document}